# NOVEL COMPONENT-BASED DEVELOPMENT MODEL FOR SIP-BASED MOBILE APPLICATION

Ahmed Barnawi, Abdulrahman H. Altalhi, M. Rizwan Jameel Qureshi and Asif Irshad Khan

Faculty of Computing and IT, King AbdulAziz University, Jeddah, Saudi Arabia

ambarnawi@kau.edu.sa, ahaltalhi@kau.edu.sa, rmuhammd@kau.edu.sa and aikhan@kau.edu.sa

***ABSTRACT***

*Universities and Institutions these days' deals with issues related to with assessment of large number of students. Various evaluation methods have been adopted by examiners in different institutions to examining the ability of an individual, starting from manual means of using paper and pencil to electronic, from oral to written, practical to theoretical and many others.*

*There is a need to expedite the process of examination in order to meet the increasing enrolment of students at the universities and institutes. Sip Based Mass Mobile Examination System (**SiBMMES**) expedites the examination process by automating various activities in an examination such as exam paper setting, Scheduling and allocating examination time and evaluation (auto-grading for objective questions) etc. SiBMMES uses the IP Multimedia Subsystem (IMS) that is an IP communications framework providing an environment for the rapid development of innovative and reusable services Session Initial Protocol (SIP) is a signalling (request-response ) protocol for this architecture and it is used for establishing sessions in an IP network, making it an ideal candidate for supporting terminal mobility in the IMS to deliver the services, with the extended services available in IMS like open APIs, common network services, Quality of Services (QoS) like multiple sessions per call, Push to Talk etc often requiring multiple types of media (including voice, video, pictures, and text). SiBMMES is an effective solution for mass education evaluation using mobile and web technology.*

*In this paper, a novel hybrid component based development (CBD) model is proposed for SiBMMES. A Component based Hybrid Model is selected to the fact that IMS takes the concept of layered architecture one step further by defining a horizontal architecture where service enablers and common functions can be reused for multiple applications. This novel model tackle a new domain for IT professionals, its ideal to start developing services as a small increments using the component and then adding new functionalities and improvements over time. The proposed novel model will be highly customizable for any university who acquired to adopt similar IMS based mass examination system. This model will be compatible with the traditional question-answer style examination and could meet the requests of the mass examination, such as university entrance exam, International Computer Driving License exam, etc.*

# 1. INTRODUCTION

Increasingly, academics are dealing with issues associated with assessment in large classes, arising from factors like high enrolment of students. Assessment of students requires a lot of effort, consumes enormous time and resources, this results in high time pressure on academics to search for alternative assessment methods that can guarantee fast assessment [1].

Setting of test paper is one of the biggest problems even with objective choice questions as it is a repetitive job. An individual has to set the test paper for a particular course. The next time when some other person sets the paper for the same course that needs rework.

There is a need to expedite the process of examination in order to meet the increasing enrolment of students at the universities and institutes [1]. Sip Based Mass Mobile Examination System (SiBMMES) expedites the examination process by automating various activities in an examination such as exam paper setting, Scheduling and allocating examination time and evaluation etc.

The SiBMMES will assess to students by conducting online/mobile based objective exam. This will be highly customizable for any university who acquired to adopt similar IMS based examination system and Faculties to create their own dashboard (create set of questions, creates groups, adds related students into the groups, schedule exams, etc). Further the exams will be associated with specific groups so that only associated students can appear for the test, result will be notified to the student either through SMS/email as shown in the underneath Figure 1.

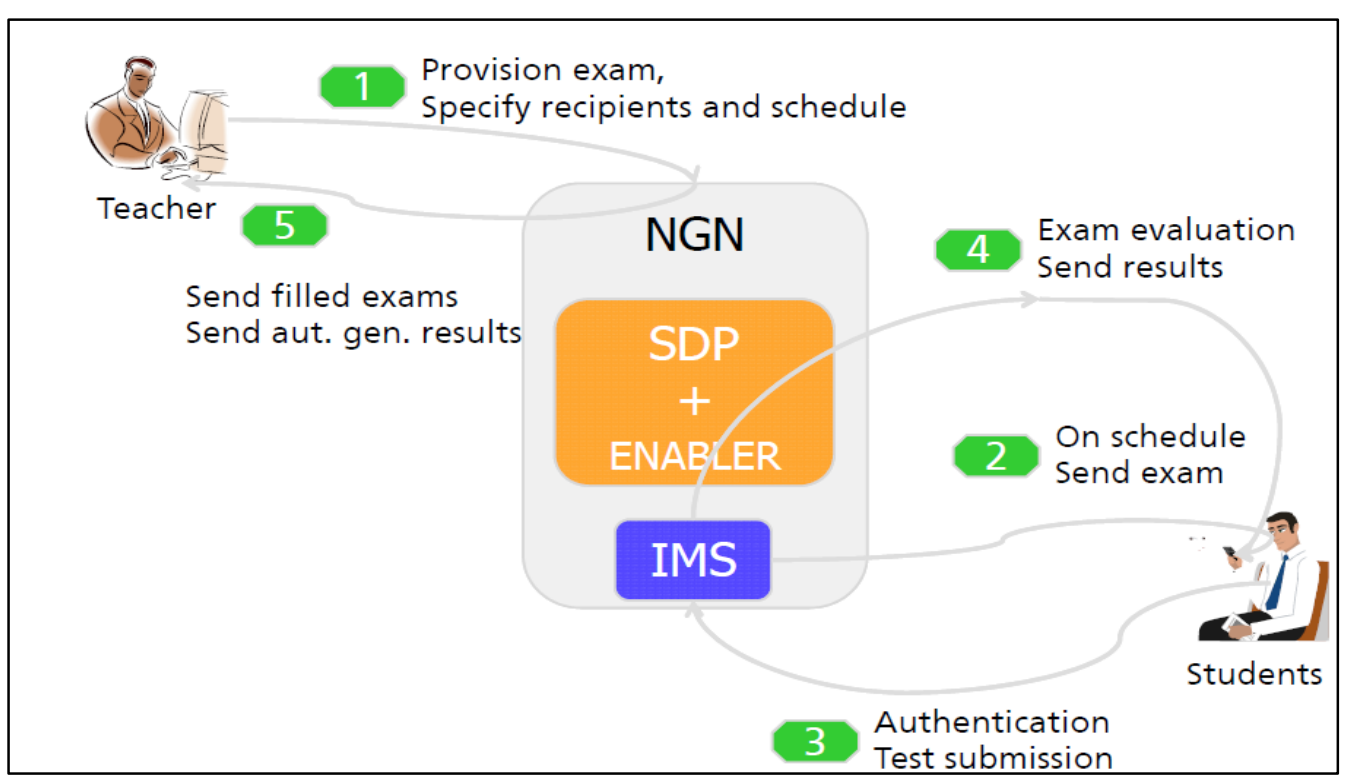

Figure 1. IMS-based mobile exam scemniro [2]

## 1.1 IP Based Multimedia Subsystem

SiBMMES uses the (IP based Multimedia subsystem) IMS platform to deliver the services. The purpose of IMS is to provide the Internet service to the user anywhere and anytime through the mobile phone technology. IMS is designed with building blocks, giving telecommunication operators the ability to deliver new services in a more flexible way. The introduction of IMS networks brings a network-agnostic server delivery model to the server providers. They are able to deliver converged services, in combination of messaging, video, data and etc, regardless of the type of the network, yet achieve a better quality of service (QoS). IMS network architecture has three main layers, which are transport layer, control layer and service layer. The separation of layers makes it easy to standardize the interfaces and interconnect the systems [3].

### 1.2 Component Based Software Development

SiBMMES software development model is based on Component Based Software Development, One of the main principles of computer science, divide and conquer the bigger problem into smaller chunks to solve it, fits into component based development. The aim is to build large computer systems from small pieces called a component that has already been built instead of building complete system from scratch.

Reuse of software components concept has been taken from manufacturing industry and civil engineering field [4]. Manufacturing of vehicles from parts and construction of buildings from bricks are the examples. Car manufacturers would have not been so successful if they had not used standardized parts/components. Software companies have used the same concept to develop software in standardized parts/components. Software components are shipped with the libraries available with software. Microsoft Corporation and Sun Microsystems are two major software-providing organizations. These companies have provided components with their software to market themselves successful and their tools are widely used in software industry. Most of the offered tools provide an IDE (Integrated Development Environment). IDE provides an environment in which components are available in the toolbox or in the reference library like a car assembling plant. We do not need to develop the components during the assembling of the car but they are there and we timely assemble them. Similarly in IDE, the standard components such as text box, label box and command button are available in the toolbox and we just integrate and use them [5].

A number of attempts, [6, 7], have been made to propose CBD models in the last few years. However, there are a few issues which are still unaddressed and there exists possibilities of improvement in the proposed CBD models. Therefore the research problem is:
Question: How to propose an improved systematic component-based development process model for development of mobile supported mass examination system to be compatible with the IMS-based environment?

The paper is organized as: **section 2** describes related work of IMS architecture and CBD Models along with comparison of the existing CBD Process Models, **section 3** Motivation for the Proposed CBD Model , and **section 4** describes proposed novel CBD Process Model.

## 2. RELATED WORK

There are many research bodies/ organizations on developing better ways to manage exams systems and assessment. Magdi Z. Rashad and his colleagues [8], they looked Web-based Exam Management Systems (EMS) as an effective solution for mass education evaluation. They proposed a web based system to conduct online exam activities such as registration, online exam, auto grading etc. the authors concluded that the presented system saves instructors from suffering and tedious of grading works, further in the paper the authors pointed out that 94 % of the students like the user interface and 85 % agree that the system is usable.

Web-based examination system is an effective tool for mass education evaluation [9]. As per the paper it is a novel online examination system based on a Browser/Server framework which performs the exam and auto-grading for objective questions and operating questions, such as programming, operating Microsoft Windows, editing Microsoft Word , Excel and PowerPoint, etc. It has been successfully applied to the distance evaluation of basic operating skills of computer science, such as the course of computer skills in Universities and the nationwide examination for the high school graduates in Zhejiang Province, China.

## 2.1 IMS Architecture

The 3GPP defines IMS as an architecture framework for delivering multimedia services for both wireless and fixed access technologies based on the Internet Protocol (IP) [2]. SIP has emerged as the vital technology for controlling communication in IP-based Networks. The IMS platform is based on layered architecture design as shown in Figure 2.

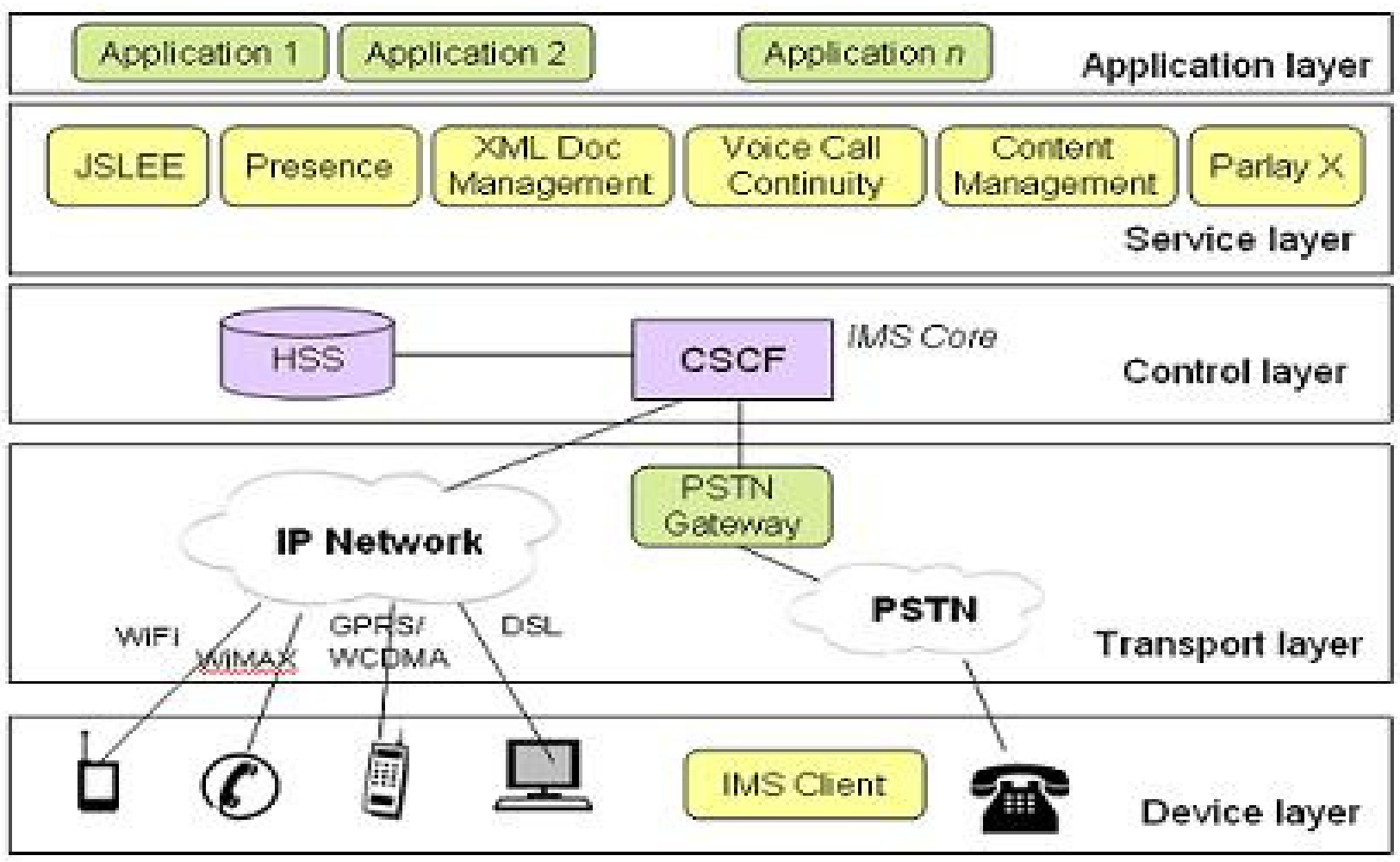

Figure 2. IMS architectural diagram Source [10]

The device layer refers to the IMS client applications and devices while service layer is manly use to provide added functionality to IMS core network.

### 2.1.1 Main layers of IMS Network

The IMS core network consists of mainly Transport layer, Control layer and Application layer.
**Transport layer-** This layer is designed to provide network access. Initiation and termination of SIP sessions is done by this layer. IMS devices connect to the IP network in the transport layer using different technique including Cable, WiFi, DSL, modem, SIP, GPRS, and WCDMA etc.
**Control layer-** This is the middle layer mainly contains Call Session Control Function (CSCF) and Home Subscriber Server (HSS). Routing and generating billing details for the use of the network are the main responsibility of this layer, further role of CSCF and HSS is mentioned in Key components section.[3]

**Application layer-** This layer allows service providers to offers different multimedia services to its users. The application servers are in charge of hosting and executing the services and provides interface against the control layers using the SIP protocol. There may be several application servers providing different multimedia services. Presence server, Instant Messaging Server, Group List Management Server etc are some core application servers.[3]

There are several internationally recognized telecommunication and Internet standardization bodies responsible for the standardization of the IMS, SIP and Java technologies like The 3rd Generation Partnership Project (3GPP), European Telecommunications Standard Institute (ETSI), The Internet Engineering Task Force (IETF), Java Community Process (JCP) etc. The JCP is also involved in the standardization of Java APIs to facilitate the easy and fast deployment of IP services using SIP. The main IMS system consists of different software components inter-operating to provide services and management to the network. Following are some of the software components support for IMS.

### 2.1.2 JAVA SUPPORT FOR SIP AND IMS

There are four Java Specification Requests (JSRs) for SIP targeting three different Java platforms namely Java 2 Standard Edition (J2SE), Java 2 Mobile Edition (J2ME), Java 2 Enterprise Edition (J2EE). These SIP JSRs can be listed as follows:-

- **JAIN SIP:** JAIN SIP is the standardized Java interface to the Session Initiation Protocol for desktop and server applications. JAIN SIP enables transaction stateless, transaction stateful and dialog stateful control over the protocol. It is targeted at J2SE and enables the development of standalone user agent, proxy and user registrar applications. Extensible by nature, it provides standardized interfaces that can be used to provide minimum SIP support in applications [11].

- **SIP Lite:** SIP Lite can be implemented both on J2SE and J2ME platform and provides an application environment for developers who are not SIP literate. Originally developed for J2SE, the specification is so small that it can be implemented on J2ME platform. The goal of this high-level API will be to allow application developers to create application's that have SIP as their underlying protocol without having to have an extensive knowledge of the SIP protocol. This will allow developers to rapidly create applications, such as user-agent type applications. It is suitable for midsize devices with relatively more memory and processing power than mobile devices e.g. SIP enabled phones [11].

- **SIP Servlet(JSR 116):** defines a high-level extension interface to enable SIP servers to deploy and manage SIP applications based on the servlet model, This API is defined for execution on network based SIP applications and is implemented on application servers supporting SIP with an option to support HTTP and J2EE [11].

- **SIP API for J2ME (JSR 180):** This specification defines a J2ME optional package that enables SIP support for limited resource mobile devices. It is the standardized SIP interface for mobile handsets with core network functionality. Although it is specifically designed for Connected Limited Device Configuration profile (CLDC), it can also be used with Connection Device Configuration (CDC) profile. Client devices have to support SIP for Rel. 5.0 of the Universal Mobile Telecommunications System (UMTS) architecture. The API is integrated with the Generic Connection Framework (GCF) of J2ME which keeps the look and feel of the HTTP API [11].

## 2.2 Component Based Software Development

According to Lycett and Giaglis [12] the evaluation of information systems in terms of reuse is extremely difficult. They are of the view that all development approaches have a danger and risk to reuse the existing components.

The main reuse risks can be avoided by following ways [12]

- Integrate business driven evaluation at an early stage during selection of components. It will reduce effort of evaluation and selection.
- Evaluation and selection should be an ongoing process.

Lycett and Giaglis [12] discussed Discount Cash Flows (DCF), Net Present Value (NPV), Return on Investment (ROI), Internal Rate of Return (IRR), SESAME and cost benefit analysis sheet (CBA) but without providing any facts and figures. They [12] did not provide comprehensive evaluation criteria to evaluate and integrate the previously developed components for reuse. They [12] proposed a content, context and process (CCP) analysis to evaluate a component for reuse. CCP analysis is not practical from implementation point of view because it is very subjective for selection and evaluation of a component for reuse. Constructive Cost Model II proposed by Barry Boehm [Boehm et al. (2000)] is far better to calculate cost of new systems to be developed by new and reusable components.

According to de Jonge [13], one of the reuse requirements is to develop independent components that must be integrateable. He [13] proposed techniques to integrate reusable independent components within one system and across systems. The proposed source tree composition technique integrates;

- core modules of components;
- package development that permits development of more than one component concurrently according to the scope of different software systems.

Package based software development is a popular research area [13]. The author concentrates on source tree composition technique for software configuration management of more than one system. This is a problem in terms of management of reusable components in a repository. He [13] did not discuss crosscutting development of components and their integration in multiple systems.

The extent of software reuse depends upon the reuse strategy followed [14]. Marcus et al. [14] proposed a set of six dimensions to support the reuse practices after conducting a survey during which data was gathered from 71 software development groups. These dimensions were planning and improvement, formalized process, management support, project similarity, object technology and common architecture. On the basis of dimensions, they [14] discovered five reuse strategies that are practiced in software development groups. The five strategies were ad-hoc reuse with high reuse potential, uncoordinated reuse attempt with low reuse potential, uncoordinated reuse attempt with high reuse potential, systematic reuse with low management support and systematic reuse with high management support. Main objective of their research is to classify reuse strategies so that development groups can get benefit and achieve success to complete software projects. The authors [23] support the last strategy i.e., systematic reuse with high management support, but this scheme needs highly detailed analysis. This analysis consists of gathering of data from various software houses working in different geographical locations, before reaching a conclusion.

Selection of reusable components is important to improve productivity of a component-based software Jihyun et al. [15]. This research proposed a Component Repository for facilitating Enterprise Java Beans (EJB) Component Reuse (CRECOR) to store and manages the reusable components. Working on the reusable components with repository (software) has many benefits. For example,

- Specification viewing
- Adapting
- Testing
- Deploying

The repository proposed by the authors Jihyun et al. [15] does not have a version control and change control functions to manage different versions of reusable components. Version and change control are important functions of a repository to manage and update different versions of a component.

Haddad [16] is of the view that software organizations have to invest huge sums of money to start successful reuse methodology and it's a barrier for them. The author [16] believes that core of reuse is source code. According to an estimate mentioned by the author, "domain specific components represent up to 65% of the application size. One approach for effective reuse practices focuses on domain specific components". The author proposed an integrated approach for component-based development to support domain specific components. An integrated approach is a collection of reusable components in a development environment. The author also discusses the concept of interface to describe a wrapper interface mechanism. The wrapper interface mechanism can be used to manage and control the interface between or among integrated collection of reusable components. The objective of this research is to develop benchmarks for software organizations so that these begin reuse practices by emphasizing mainly on programming effort and not on management and operational issues. Focus of author's research is development of domain specific reusable components and not construction of reusable components of different domains of concern [16]. This problem can be handled through software engineering for adaptive and self-managing systems. Geisterfer and Ghosh [17] proposed few recommendations for component selection without validation.

The research by Arndt, J., et al. [18] reveals that there are two traditional approaches to construct software systems i.e., customization and use of standard components libraries. A composition of customized and component libraries domains approaches has been proposed to achieve benefits of both. There are many factors to resist practice of this new domain. A change process is explained to support composition approach by providing a logical solution. A concept is also introduced for mindful innovation to show that how modular development can avail the benefits of domain change. The advantages of Component Based Software Development (CBSD) domain are also discussed that are already described in many papers. This work explains this concept theoretically. However, this research is conducted to support the practice of CBSD domain [18]. CBSD acceptance process is not considered by the authors.

Khemakhem et al. [30] presented SEC (search engine for component-based development). SEC helps developers to identify and assemble components using a repository. It does not use ontology-based description technique to retrieve components. An advanced version of SEC is presented by the authors in [Khemakhem et al. (2007)]. This version uses ontology-based technique to identify and retrieve components. The searching engine is deficient in wrapping complete set of ontologies. SEC has not been tested for a variety of CASE tools. A model is proposed by the authors [Ha and Lee (2006)] to merge Ontology Web Language (OWL) based framework and web services for CBD. It manages components using ontology based technique. It does not handle automatic configuration and generation of ontologies to search components. Jiang Guo [19] discusses various integration issues in the current enterprise resource planning (ERP) systems. The author suggests that most of the integration issues are resolved by using category theory. This research objective is to propose a framework for component dependencies modeling technique [19]. The proposed framework requires further evaluation through case studies.

The proposed model [21] integrates components using method type collection. It facilitates software engineering team in component qualification by using signatures. The connection model is based on Java and it does not support other architectures such as Microsoft. The concept needs further validation for large and complex applications [21]. A comparative analysis of the existing process models has been presented by Crnkovic and his colleagues [28]. The objective is to describe different phases of component-based system life cycle. The authors have partitioned component-based development processes into system development and component development. They have not introduced any new concept. CBSE is already divided into two domains which are CBD and domain engineering [22].

Teiniker et al. [29] propose a hybrid development process for component-based software systems. It consists of model-driven system development, test-driven component development and quality of service driven system deployment processes. A case study validates the proposed model for a reengineering project. The drawbacks of the proposed processes are [29]:

- use of unified modeling language (UML) to integrate different models;
- Rational Rose tool generates code using the model, if model is changed complete code is generated again.

### 2.2.1 A Comparison of the Existing CBD Process Models

A software-cycle model was proposed for reuse and reengineering [20]. The proposed model suggested five stages to reuse a component.

- Analysis of existing programs to sort components to be reused.
- Reengineering to eliminate domain specific troubles.
- Saving reusable components in the repository.
- Construction of independent status components with a reuse approach to store in a repository.
- Reuse components to develop new programs.

It was not a complete process model for CBD but main focus of their research was on reuse activities [20]. The drawback of this CBD model was removed by proposing an improved CBD model [23].

Main phases of improved CBD model were 'Component Analysis', 'Architectural Design', 'Component Brokerage', 'Component Production' and 'Component Integration' phases [23]. The improved CBD model was a modification of Waterfall model integrated with these phases. There were some limitations in the improved CBD model. The improved CBD model was not suitable for commercial applications because of the verification of phases repeatedly. It was more time and cost consuming process model and suitable only when comprehensive or stable requirements were available to software engineering team. A CBD model was proposed in 2001 to remove the limitations of improved CBD model [21].

A software life cycle model was anticipated to support CBD using object oriented (OO) construction [21]. According to authors, the main phases of CBD model were 'Domain Engineering', 'System Analysis', Design and Implementation. The major problem of their model was the selection of reusable components during the design phase. The selection of reusable components should be during the analysis phase. Therefore, analyst could estimate the cost, schedule and effort required to develop and integrate the components. The efficiency of existing CBD models was improved in 2003 [25].

Poorly gathered SW requirements could fail a SW project [25]. It was because of natural drawback in requirements determination methods. An approach was proposed to construct SW by categorizing components in a knowledge base. Existing components were recognized, chosen and integrated in a newly developed SW by using the knowledge base. Different classification schemes to reuse artefacts had been discussed as well. These were enumerated, keyword, faceted and hypertext. The objective of this paper was to ease requirements gathering using knowledge base but the paper lacked in suggesting a comprehensive process model for CBD. An attempt was made in 2004 to remove the limitations of existing CBD models [24].

Table 1- A Comparison of Existing CBD Process Models

| **Existing CBD Process Models** | **Main Drawbacks** |
|---|---|
| The Software-Cycle Model for Reengineering and Reuse [20] | The model proposed by authors was not a complete process model for CBD but core focus is on reuse activities. The model also lacked in domain engineering activities. |
| A Component-Based Software Development Model [23] | The proposed model was time consuming, costly model and suitable only when comprehensive or stable requirements were provided. |
| Component-Based Software Process [27] | The major problem of this model was the selection of reusable components during design phase instead of analysis phase resulting in poor estimation of cost, schedule and effort to develop and integrate components. |
| An Assessment Model for Requirements Identification in Component-Based Software Development [25] | The model lacked in suggesting a comprehensive process model for CBD. |
| Distributed Component-Based Software Development: An Incremental Approach [24] | A clear-cut process model was not proposed and CASE tool specific for the development of CBD projects. |
| A Service Model for Component-Based Development [26] | Repository was not used in this model for the development of CBD projects which was an essential requirement for reuse. |

An incremental method was presented for distributed CBD [24]. It was based on two phases. The first phase composed of gathering requirements of the problem domain and construction of employable components in object-oriented (OO) language. These employable components were stored in a repository. Software engineers looked up MVCASE tool to select the necessary components to develop the SW system in the second phase. There were two main weaknesses of this model. It was not a transparent model and a use of a specific CASE tool was the requirement of this process model. The efficiency of existing CBD models was improved [26].

The four-stage component-based development process model was very complex for implementation [26]. The core objective of their research was to integrate off-the-shelf components with the newly developed components rather than in house development. Repository had not been used here. A comparison of component-based development process models is shown in Table 1.

Table 1 lists certain unaddressed issues and there is a room for improvement in the existing CBD models. Therefore, the research problem focuses on the drawbacks of existing CBD models to improve them.

The main research problem (based on related work) is "How to propose a novel hybrid component development model for SIP based IMS Mobile Application? ".

## 3. MOTIVATION FOR THE PROPOSED CBD MODEL

The CBD model proposed by authors [John and Victor (1991)] was not that comprehensive. In fact its main focus was on reuse activities and it lacked in domain engineering activities. The domain engineering activities are vital to populate the repository to reuse components in current and future software projects. The said limitation in the CBD model (of John and Victor) motivates the author to propose a new CBD model that includes both CBD and domain engineering activities.

Ning [23] proposed another CBD model that was a modified version of Waterfall model. It was a time consuming and costly model due to verification of phases like Waterfall model. This model emphasized on freezing the requirements. That is why; it was not suitable for commercial projects where requirements of the project development are dealt dynamically. The model was suitable only when comprehensive or stable requirements were provided from the very beginning. The shortcomings of CBD model (23) become source of inspiration to propose a new CBD model that will deal with run time requirements dynamically to meet the needs of commercial software projects.

An incremental model was proposed for CBD by the author [24]. Analysis and construction were only two phases in the incremental model. It was neither that comprehensive nor properly covering system development life cycle phases for CBD software projects. The usage of specific case tool to adopt the model was another limitation. The limitations of their model encourage the author to propose a new CBD model. The new CBD model will have relatively more comprehensive system development life cycle phases for both CBD and domain engineering activities. Moreover, there will be no constraint to use any specific case tool while using the proposed model to develop CBD software projects.

A four-stage component-based development process model was proposed by the authors [26]. They introduced a new technique to integrate off-the-shelf components with the newly developed components in the then existing CBD process models. However, they paid less rather no attention to the integration of internally developed components. The same motivates the author to propose a new CBD model.

The author [21] used repository on the design phase. The design phase is not that appropriate phase to use repository. This is because; using repository at analysis phase helps a software engineering team to identify new and reusable components. Moreover, it has many benefits such as; initiation of CBD and domain engineering activities and more accurate estimation of cost, schedule and effort to develop and integrate components in a CBD project. The proposed CBD model will use repository at analysis phase to avail the above mentioned benefits.

It is with this logic and background in view that the author feels motivated to propose a new CBD model. The same is accomplished by proposing a new CBD model as a solution to the research problem as follows.

# 4. THE PROPOSED CBD MODEL FOR MOBILE APPLICATION

A new component-based development (CBD) model has been proposed for an IMS-based mass mobile examination system as a solution for the research problem. A CBD model is a process model that provides a framework to develop software from previously developed components. The main phases of the improved CBD are 'Project Planning', 'Analysis', 'Adaptation, Engineering & Integration' and 'Testing'. An improved CBD model to be proposed is shown, in figure 3.

**Project Planning Phase**- A customer is communicated at the start of the project to gather basic requirements. Initial use cases are developed at this stage to prepare project specification or proposal document is prepared during the 'Project Planning' phase. Project specification or proposal document is composed of feasibility and risk assessments that are performed to prepare a cost benefits analysis (CBA) sheet. CBA sheet helps to estimate whether the software project is feasible for the customer or not.

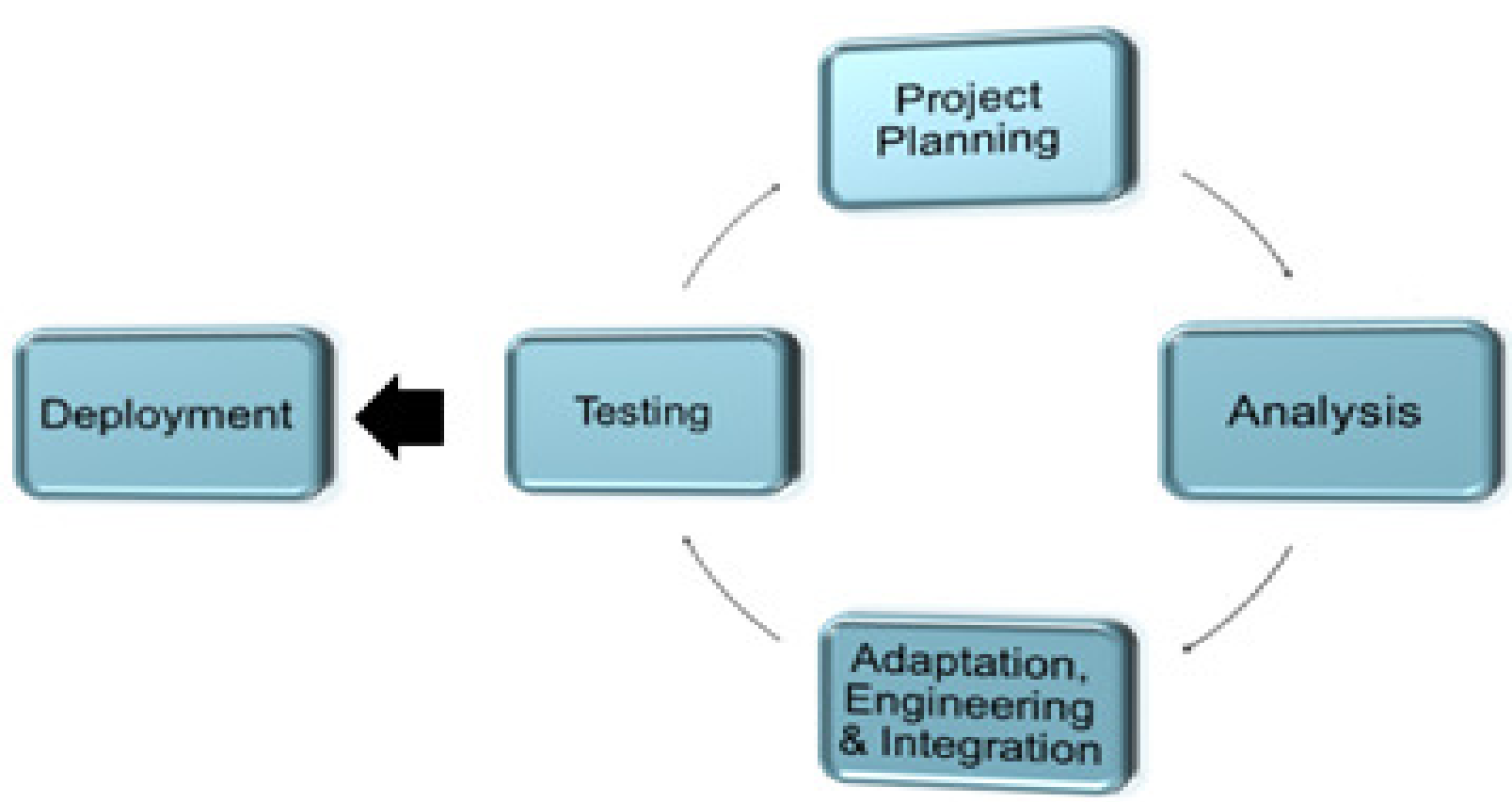

Figure 3 -The Proposed New Component-Based Development Process Model

**'Analysis' Phase**- Analysis phase is initiated if the customer approves the proposal. This is the phase where an analyst gathers the detailed requirements of the system to be developed. A domain analysis is performed to find a suitable architecture for the application to be developed [21]. An architectural model of application is developed that enables a software engineer to:

- evaluates efficiency of design;
- judge options of design;
- minimize potential threats coupled with software development [23].

'Analysis, Selection & Risk Management' is the phase where an analyst tries to identify and select those components that can be reused from the components repository. The selection of reusable components is important to improve productivity of component-based software development. Risks about new and existing components are also evaluated and managed. Software engineering methods are applied to develop new components for those requirements which cannot be fulfilled from already developed components.

Reusable components need qualification, adaptation and composition. Component qualification makes sure that the selected component will:

- execute the desired functionality;
- integrate easily into the structural design of new application;
- demonstrate the quality attributes (e.g., reliability, performance, usability) [21].

The properties, behaviour and relationships among components are identified. Core objective of this phase is to reuse components as maximum as possible, rather than reinventing the wheel. It will also improve productivity and efficiency of software engineers.

**'Adaptation, Engineering & Integration' Phase**- The reusable components are customized according to the requirements of the new system to be developed and tested. The next part is integration and testing again. A common component adaptation technique called component wrapping is used if programmer is using black box components [23].

**Testing Phase**- The new components are designed, developed and tested on unit basis. Integration and system tests of the newly developed and of the reused components are performed. A customer is requested to evaluate and verify software, whether it meets his/her requirements or not during the testing phase. The software is ready to deploy at customer site.

## 5. CONCLUSIONS

An ongoing research is presented in this paper by putting together a platform as a testbed for NGN application development. We propose a novel component based development model used for this SIP based mobile applications. The proposed model used as a framework for general purpose application development over the testbed. The objectives of IMS based Mobile examination System approaches are explained with reasons and advantages identified. Main components of IMS service architecture with their roles are also described. The approach leads to a highly modular and extensible integrated system. Future work is to validate novel component based development model using a case study of IMS testbed. Multi-tier applications architecture (client, web, and business) is adopted, as per the needs of case study i.e., MVC design pattern.

## ACKNOWLEDGEMENT

The authors would like to thank **King Abdulaziz City for Science and Technology (KACST),** Saudi Arabia for funding this ongoing research project.

## Author Bibliography

**Dr. Ahmed Barnawi** received his BSc in Electrical Engineering from King Abdul-Aziz University in 2000, his degree in Communication Engineering from University of Manchester Institute of Science and Technology (UMIST) in 2002, and his PhD degree in Mobile Communications from Bradford University in 2006. Currently, Dr. Barnawi is an Assistant Professor at the Department of Computer Science, King Abdul-Aziz University, Jeddah, Saudi Arabia. His current research interests include Mobile Next Generation Network, Cognitive Radio and Wireless Ad hoc and Sensor Networks.

**Dr. Abdulrahman Altalhi** is an assistant professor of Information Technology at King Abdul-Aziz University. He has obtained his Ph.D. in Engineering and Applied Sciences (Computer Science) from the University of New Orleans on May of 2004. He served as the chairman of the IT department for two years (2007-2008). Currently, he is the Vice Dean of the College of Computing and Information Technology. His research interest include: Wireless Networks, Software Engineering, and Computing Education.

**Dr. M. Rizwan Jameel Qureshi**, is an assistant Professor at IT Department, Faculty of Computing and Information Technology, King Abdul Aziz University, Jeddah, Saudi Arabia. He has done his Ph.D. CS (Software Process Improvement) in 2009. He is in the field of teaching and research since 2001. He has published twenty two research publications and six books at national and international forums. He is teaching Software Engineering domain courses at graduate and undergraduate level from more than ten years.

**Mr. Asif Irshad Khan** received his Bachelor and Master degree in Computer Science from the Aligarh Muslim University (A.M.U), Aligarh, India in 1998 and 2001 respectively. He is presently working as a Lecturer Computer Science at the Faculty of Computing and Information Technology, King Abdul Aziz University, Jeddah, Saudi Arabia. He has more than Six years experience of teaching as lecturer to graduate and undergraduate students in different universities and worked for four years in industry before joining academia full time. His research interests include Software Engineering, Component Based Software Engineering and Reuse.